# Direct synthesis of single-crystalline bilayer graphene on various dielectric substrate

Zuoquan Tan[1,2 #], Xianqin, Xing[1,2 #], Yimei Fang[3], Le Huang[4], Shunqing Wu[5], Zhiyong Zhang[4], Le Wang[1,2], Xiangping Chen[6, *] and Shanshan Chen[1,2 *]

[1] School of Physics and Beijing Key Laboratory of Optoelectronic Functional Materials and Micro-nano Devices, Renmin University of China, Beijing 100872, China

[2] Key Laboratory of Quantum State Construction and Manipulation (Ministry of Education), Renmin University of China, Beijing 100872, China

[3] School of Science, Jimei University, Xiamen 361021, China

[4] Key Laboratory for the Physics and Chemistry of Nanodevices and Center for Carbon-based Electronics, School of Electronics, Peking University, Beijing 100871, China.

[5] Collaborative Innovation Center for Optoelectronic Semiconductors and Efficient Devices, Department of Physics, Xiamen University, Xiamen 361005, China

[6] Tsinghua Shenzhen International Graduate School, Tsinghua University, Shenzhen, 518055, China

[#] Z. Tan and X. Xing contributed equally to this work.

*Correspondence: schen@ruc.edu.cn and cxp0826@sz.tsinghua.edu.cn

## Abstract

The growth of high-quality Bernal-stacked bilayer graphene (BLG) directly on dielectric substrates is crucial for electronic and optoelectronic applications, but there are still challenges such as poor quality, uncontrollable thickness and polycrystalline films. In this work, a novel method to grow high-quality and single-crystalline BLG directly on various dielectric substrates ($SiO_2$ /Si, sapphire, and quartz) was demonstrated. Single-crystalline monolayer graphene was applied as a seeding layer to facilitate the homo-epitaxial synthesis of single-crystalline BLG directly on insulating substrates. The Cu nano-powders (Cu NPs) with nanostructure and high surface-area were used as the remote catalysis to provide long-lasting catalytic activity during the graphene growth. The TEM results confirm the single-crystalline nature of the BLG domains, which validates the superiority of the homo-epitaxial growth technique. The as-grown BLG show comparable quality with the CVD-grown BLG on metal surface.

Field-effect transistors directly fabricated on the as-grown BLG/$SiO_2$/Si showed a room temperature carrier mobility as high as 2297 cm$^{2}$ V$^{-1}$ s$^{-1}$.

## 1. Introduction

Bernal-stacked bilayer graphene (BLG) features appealing physical properties, including the tunable electronic bandgap[1-3] and the intrinsic valley degree of freedom[4-6], making it a candidate of interest for electronic, optoelectronics and quantum devices[7, 8]. Conventional chemical vapor deposition (CVD) on metallic catalysts-including Cu[9-11], Ni[12] and Cu-Ni[13-15] alloys-has yielded large-area, high-quality AB-stacked graphene. In contrast, insulating substrates offer limited catalytic activity for precursor dissociation and impose high surface-diffusion barriers for carbon species (on the order of several eV)[16]. Consequently, direct growth on insulators typically relies on elevated temperatures[17] and prolonged growth times[18], often resulting in non-uniform thickness, high defect densities, and abundant grain boundaries[18, 19], which constrain film quality and limit device applicability.

Substantial progress has nevertheless been employed to achieve the direct growth of graphene on various dielectric substrates, including but not limited to, $SiO_2$/Si[20-25], Si[26, 27], $Si_3N_4$[18, 28], sapphire[29-31], quartz[32-35], glass[36] and h-BN[37, 38]. These advances have followed two complementary directions. The first is energy regulation—via tailored thermal fields[21, 31], plasma assistance[27, 29, 38, 39], or free molecular flow conditions[34]—to enhance precursor activation and favor edge-limited lateral propagation. However, the as-grown graphene exhibits small grain sizes due to restricted lateral growth on insulating substrates, compromised crystallinity evidenced by prominent D-band peaks in Raman spectra, and uncontrolled layer thickness variation during synthesis. The second is catalytic assistance-via deposited thin metal catalyst[20, 25, 28, 30, 32, 40], metal vapor-assisted growth[22-24, 26, 33] or non-metallic catalyst (such as $CO_2$)-assisted growth[35-37, 41] to promote nucleation and lateral growth. Although an interfacial metal catalyst thin layer can markedly improve the quality of as-grown graphene, it re-introduces the long-standing issues of metal removal and metallic-residue contamination. Delivering metal vapor remotely or turning to non-metallic catalysts can suppress dopant incorporation and avoid post-growth etching, yet neither strategy overcomes the intrinsically poor nucleation on bare insulators, and the resulting films remain highly defective. Moreover, precise layer-number control of graphene insulating substrates remains an open challenge.

Here, we report a substrate-agnostic homoepitaxial strategy for the direct synthesis of single-crystalline BLG on dielectric substrates. By employing single-crystalline monolayer graphene (SLG) as a template layer on SiO□/Si, sapphire, and quartz, we precisely control the in-plane crystallographic registry. Copper nanopowders (Cu NPs) enable remote activation and flux regulation of CH□ decomposition, allowing fine-tuned supply of reactive carbon species. Through optimization of the temperature–pressure–gas-composition window, the process favors second-layer nucleation and lateral growth, yielding BLG with orientation-aligned crystalline inheritance and deterministic layer number control. Comprehensive characterization via Raman spectroscopy and transmission electron microscopy (TEM) confirms the Bernal-stacked structure with exceptional quality, evidenced by negligible D-band intensity and sharp G-peak profiles. Field-effect transistors (FET) fabricated directly on graphene/$SiO_2$/Si substrate had a carrier mobility as high as 2297 $cm^2 V^{-1} s^{-1}$.

## 2. Experimental methods

### *2.1 Preparation of Single-Crystalline SLG seeding layer*

Large area single-crystal graphene were firstly grown on electropolished Cu foil (25 µm thick, Alfa Aesar, stock No.46365) using a Cu enclosure configuration.[42] For typical growth, the Cu enclosure underwent vacuum annealing at 1030 $^{o}$C for 10 min (partial pressure: 2 mTorr), followed by graphene deposition in 1 sccm $CH_4$ and 50 sccm $H_2$ for 3h (partial pressure: 586 mTorr). After growth, the system was cooled down rapidly to room temperature under the same gaseous atmosphere. The resulting graphene films on the inner surface of Cu enclosure were transferred onto various dielectric substrates (e.g. 300nm $SiO_2$/Si, sapphire and quartz) using a PMMA-assisted wet transfer method[42].

### *2.2 Synthesis of Single-crystalline BLG*

The synthesis of single-crystal BLG on single-crystal SLG was carried out in a low-pressure CVD system. Cu NPs were employed as the floating catalysts, strategically positioned in the upper gas stream to enable remote activation of hydrocarbon decomposition. As depicted in Figure 1a, a small quartz tube sealed at one end was loaded with Cu NPs and positioned with its open end facing downstream, ~8 cm from the Si substrate. This distance, optimized to balance catalytic efficacy against particle contamination, furnishes the maximum enhancement of growth without depositing Cu nanocrystals on the wafer. The insulating substrates with pre-transferred SLG were loaded into the center of the quartz tube furnace. Thermal annealing at 1000 °C was performed for 2h min under a 10 sccm $H_2$ (partial pressure of 120 mTorr) to remove transfer-induced residues and achieve pristine graphene seeding surfaces. $CH_4$ flow rates were systematically varied (5, 10, 15, and 20 sccm) to modulate carbon flux. Following 1 h exposure to the $CH_4$ /$H_2$ mixture, the system underwent rapid quenching to room temperature to terminate growth.

*2.3 Characterizations*

Scanning electron microscopy (SEM) imaging was conducted using a Zeiss Sigma scanning electron microscope. Transmission electron microscopy (TEM) was performed on an FEI TECNAI G Ƭ20 transmission electron microscope operated at 200 kV. Raman measurements were recorded at ambient temperature using a micro Raman spectrometer (Alpha 300, WiTec) with a 488 nm laser.

*2.4 Device Fabrication and Electrical Measurements*

The electrical properties of the single-crystal BLG were characterized by fabricating dual-gate graphene FET devices directly on the as-grown graphene on 300 nm $SiO_2$ /Si substrates. The device fabrication composed of four processes based on electron-beam lithography (EBL). First, the graphene channel was patterned with dimensions L/W = 9 μm/5 μm, followed by etching excess graphene using oxygen plasma (Power = 100 W, 40 s). Second, a 3 nm thick yttrium strip was deposited cross the channel. After exposure in air at 240 °C for 30 min, Yttrium oxide ($Y_2O_3$) (5 nm) formed above the strip. Third, a 20 nm Hafnium oxide ($HfO_2$) layer was deposited by atomic layer deposition (ALD), serving as the top gate dielectric in conjunction with $Y_2O_3$. Finally, the source/drain/gate electrodes (Ti/Au: 5/30 nm) were deposited via e-beam evaporation, while the doped silicon substrate functioned as the back gate. Electrical measurements were performed in air at room temperature using a Keithley 4200SC semiconductor parameter analyzer to characterize the FET properties.

## 3. Results and Discussion

Figures 1a and 1b schematically depict the graphene growth reactor and the corresponding deposition process on insulating substrates. Cu NPs were employed as a remote catalyst due to their nanoscale morphology and high specific surface area, which provide a stronger and more sustained catalytic activity than Cu foil. To clarify how the seeding layer influences homoepitaxial adlayer formation, we compared growth on monolayer, bilayer and trilayer graphene templates (Figure 1c-e). After 1h growth (See Experimental methods section for more details), the secondary graphene adlayer domains had nucleated on all three templates (Figure 1f-h). It's noticed that the domain size and nucleation density of the as-grown graphene varies on the different thickness of the seeding layer. As shown in Fig. 1i, the nucleation density increases as the number of graphene seeding layers rises from one to three, whereas the lateral grain-growth rate drops sharply from one to two layers and then plateaus. To gain an insightful understanding on the evolution of the growth behavior, density functional theory (DFT) calculations were carried out (see Supporting Information for more details). With increasing numbers of graphene seeding layers, the adsorption energy of the first C atom decreases rapidly, while that of the second C atom increases markedly (Fig. 1j). These results indicate that, as the seeding layer becomes thicker, lateral enlargement of existing ad-layer islands requires more energy than the formation of new nuclei, which accounts for the experimentally observed high-nucleation/low-growth regime. Furthermore, in our experiments no graphene nuclei are observed on bare $SiO_2$ /Si after 2 h; when the growth time is extended to 3 h, only micron-sized flakes with poor quality appear (See SI for more details). Taken together with the DFT results, these observations underscore the crucial role of the graphene seeding layer in promoting and directing the homo-epitaxial growth of ad-layer graphene. Specifically, a monolayer-graphene template delivers an optimal nucleation density for the adlayer while imposing only a modest energy barrier to lateral island expansion, thereby enabling reproducible growth of bilayer graphene. Moreover, because the energy barrier for lateral expansion after nucleation on monolayer graphene is lower than that for secondary nucleation on an existing bilayer, only a second graphene layer forms on the monolayer seed; further multilayer growth is suppressed. This intrinsic selectivity enables uniform, layer-precise synthesis of strictly bilayer graphene.

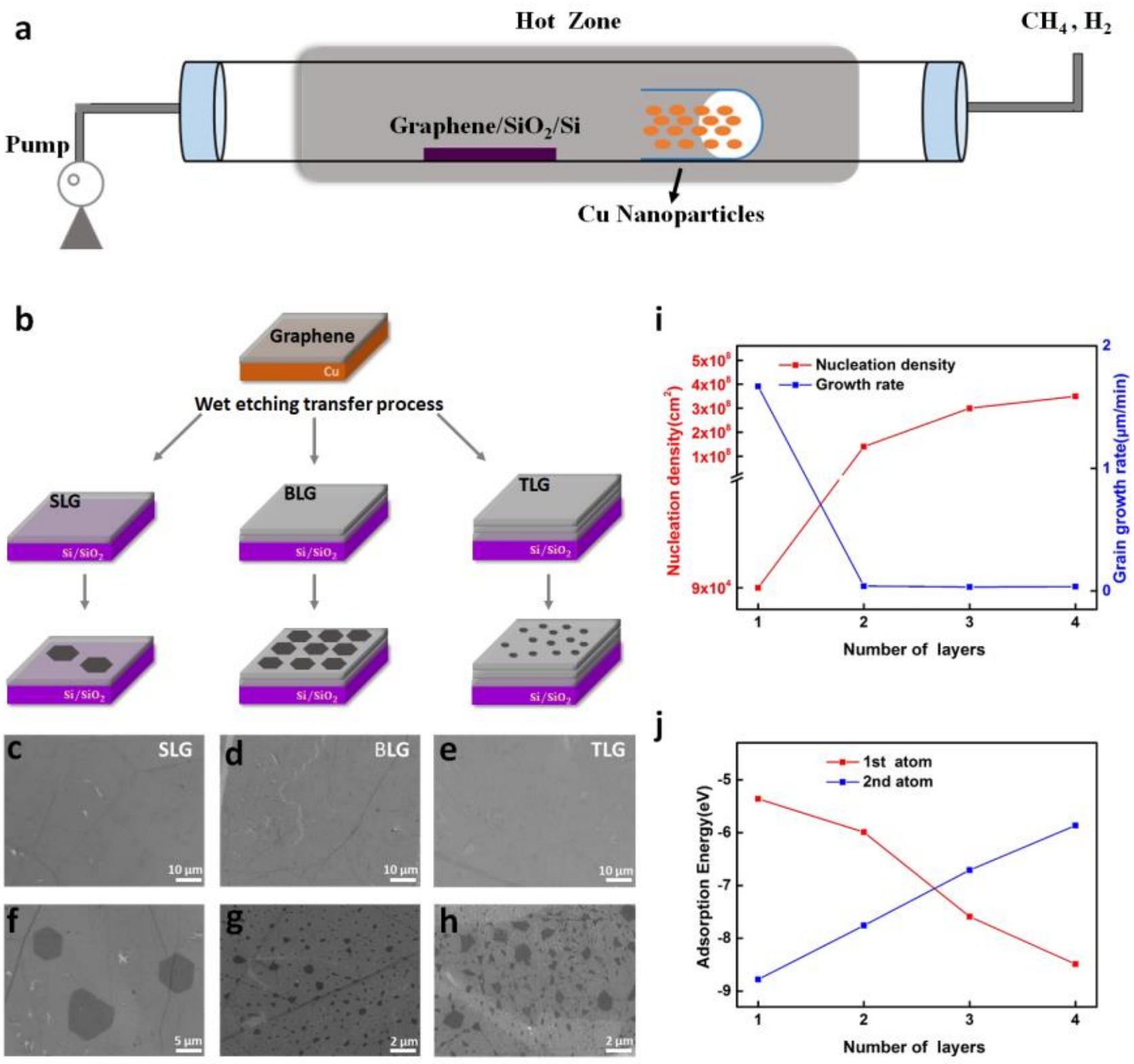

**Figure 1.** Seeding-layer-dependent homo-epitaxy of graphene on dielectric substrates. (a) Schematic of the low-pressure CVD system employing floating Cu NPs under $CH_4$ /$H_2$ . (b) Workflow: SLG growth on Cu, wet transfer onto dielectrics, followed by homoepitaxial overgrowth. (c-e) SEM images of the as-transferred SLG, BLG, and TLG seeding layers on 300 nm $SiO_2$ /Si, acquired prior to the epitaxial growth step. (f-h) SEM images after 1 h of homoepitaxial growth, showing quasi-hexagonal adlayer graphene domains on the corresponding graphene template in (c–e). (i) Nucleation density (left axis) and lateral grain growth rate (right axis) as functions of seeding layer thickness (number of graphene layers). (j) DFT-calculated adsorption energies for the first and second C atoms versus seeding layer thickness.

The key variables for Cu NPs assisted graphene-seeding growth are the annealing duration and $CH_4$ concentration. In contrast to the typical process for graphene growth, we annealed the substrates at 1000 °C in $H_2$ flow for 2h to remove transfer residues and activate growth sites. Typically, no growth would happen during the first 2 h even with $CH_4$ flow throughout the whole procedure. In addition, the effect of $CH_4$ concentration was studied by varying $CH_4$ flow while maintaining $H_2$ flow the same (10 sccm). Figure 2a-d show the SEM images of BLG with different growth gas ratio (including $CH_4$:$H_2$ = 5:10, 10:10, 15:10, and 20:10). Both the areal coverage and the nucleation density first increase and then decrease with the increasing $CH_4$ flow, yielding an inverted-V dependence (Fig. 2e). Some BLG domains even merges together to form the continuous film in Figure 2b. Raman spectra acquired from samples grown at the different $CH_4$ flows (Fig. 2f) indicate high-quality BLG (negligible D band) for $CH_4$ $\leq$ 15 sccm, whereas at 20 sccm a pronounced D band and a broadened G band emerge, consistent with reduced crystallinity[43, 44]. As a consequence, the condition of $CH_4$:$H_2$ = 10:10 can be considered as the optimized condition to obtain BLG with higher growth efficiency. Under the optimized condition ($CH_4$ :$H_2$ = 10:10), large BLG domains with hexagonal or lobed morphologies are obtained, with diagonal lengths up to 69 μm after 1 h on SLG/$SiO_2$ /Si (Fig. 2g,h)—substantially larger than the ~11 μm monolayer domains directly synthesized on $SiO_2$ /Si after 72 h[18]. Besides, Cu NPs–assisted seeding approach is further compatible with other dielectric substrates, including quartz and sapphire (See Fig. S2 in the SI for more details).

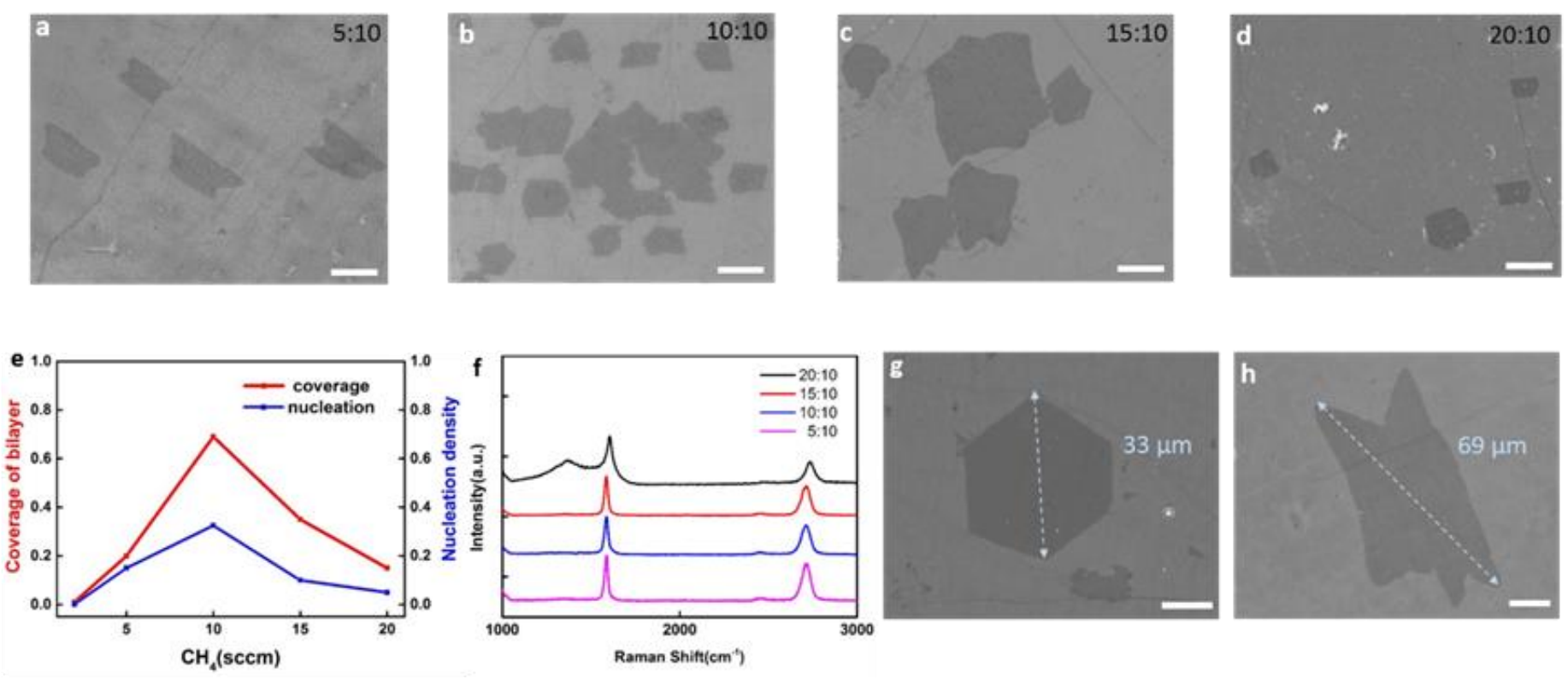

**Figure 2.** Effect of $CH_4$ supply on BLG growth on SLG/$SiO_2$ /Si substrates. (a-d) SEM images of BLG synthesized on SLG/$SiO_2$/Si substrates under identical conditions except for the $CH_4$/$H_2$ ratio for 5:10 (a), 10:10 (b), 15:10 (c), and 20:10 (d). (e) Areal coverage and nucleation density of BLG as functions of $CH_4$ flow at a fixed $H_2$ flow of 10 sccm. (f) Representative Raman spectra of the samples grown at various $CH_4$ flow. (g-h) SEM images of BLG domains with hexagonal and lobed morphologies, the largest diagonal reaches to 69 μm. Scale bar: 10 μm.

The stacking order and uniformity of the as-grown bilayer graphene were further evaluated by Raman spectroscopy. Figure 3a presents an optical micrograph of a distcrete BLG domain. Raman spectra confirm the bernal stacked nature of the as-grown BLG (Fig. 3b). Figure 3c-d show the Raman maps of the G peak (1520 – 1640 $cm^{-1}$) and the full width of half maximum (FWHM) of 2D peak (2620 – 2780 $cm^{-1}$) from the region shown in Figure 3a, respectively. The relative dark region in Figure 3c corresponds to the SLG and the bright regions corresponds to the BLG. Together with the FWHM of 2D map in Figure 3d, it is clearly that large BLG domain in whole exhibit Bernal stacking feature. Counts all of the BLG domains that obtained under the optimized growth condition, the AB-stacked bilayer graphene consists of ~ 96% of the total bilayer area (Figure 3e). Figure 3g and 3h show the $I_{2D}/I_G$ ratio of the Raman spectra used to determine the AB stacking and the histograms of 2D band FWHM values. The $I_{2D}/I_G$ ratio of the as-grown BLG is 0.95 ±0.5 and the FWHM values of the 2D band are 48.0 ± 2 $cm^{-1}$. In comparison with the 2D band FWHM values from the transferred AB-stacked BLG synthesized directly on Cu foil (Figure 3i, FWHM = 53 ± 2 $cm^{-1}$), and the sample mechanical stacked by two monolayer graphene films on the $SiO_2$/Si (Figure 3j, FWHM = 58 ± 2 $cm^{-1}$), the BLG growth on seeding layer displays a smaller FWHM which indicates an superior quality[45, 46].

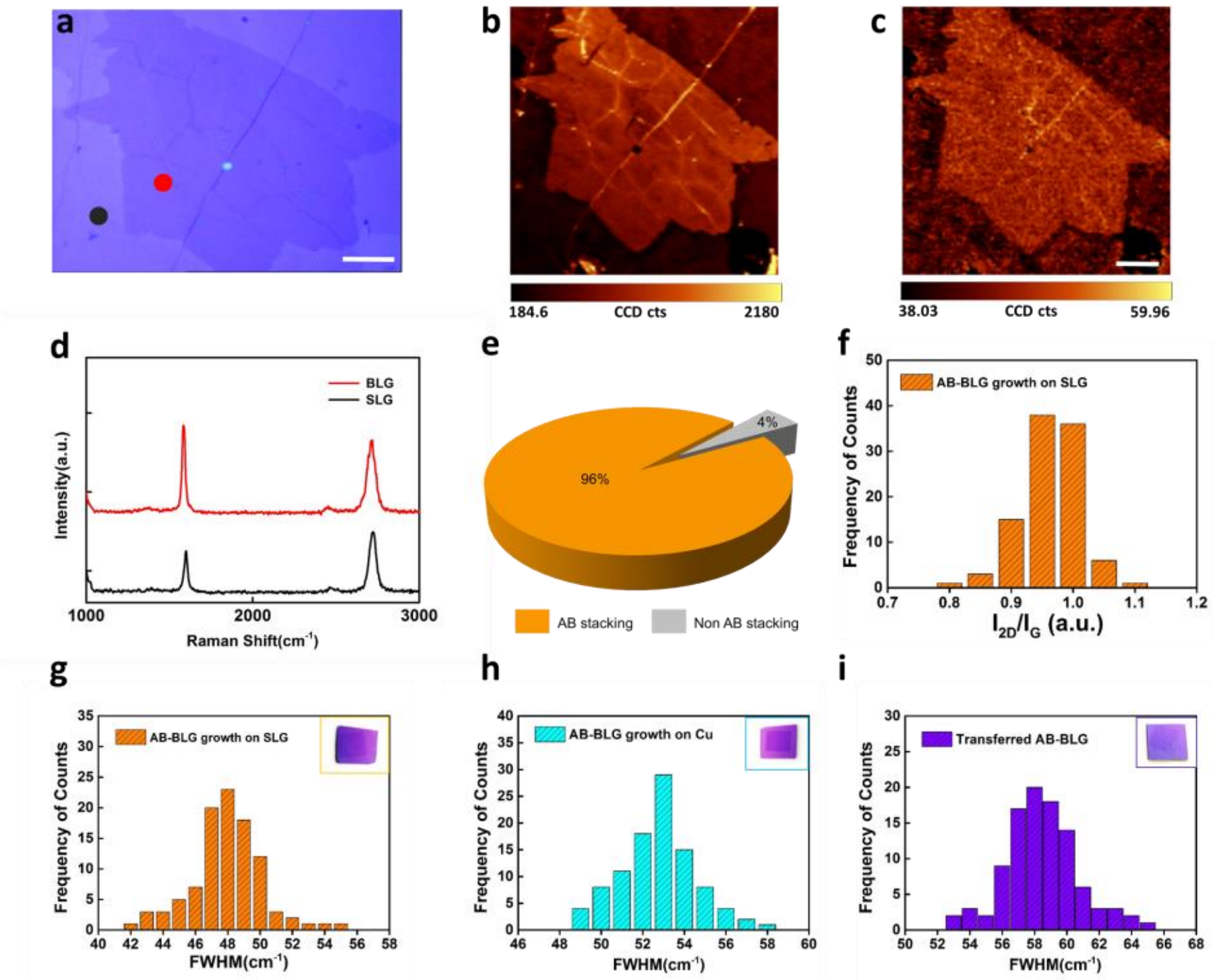

**Figure 3.** (a) Optical image of a BLG domain. (b-c) Raman map of G band intensity (1520 − 1640 $cm^{-1}$) and 2D band FWHM (2640 – 2780 $cm^{-1}$) of the same BLG grain in (a). (d) Raman spectra of the SLG and BLG areas marked in (a). (e) Stacking ratio of the BLG grown based on optimized condition. (g-i) Histograms of the Raman

spectrum 2D band FWHM values of BLG grown through the seeding technique (g), directly grown on Cu foil(h), and by transferred additional monolayer graphene onto the SLG/$SiO_2$/Si substrate (i). The insets show the corresponding photographs of the samples on 300 nm $SiO_2$/Si wafer. Scale bar: 10 μm.

The epitaxial growth nature was further evidenced by the crystalline structure of the as-grown BLG studied by transmission electron microscopy (TEM). The BLG film grown on large single crystal SLG seeding substrate was further transferred onto the TEM grid. A low-magnification SEM image is displayed in Fig. 4a, where blue and red circles mark the SLG seed and the overgrown BLG regions, respectively. Typical selected area electron diffraction (SAED) pattern of the seeding graphene in Figure 4a generate a single set of 6-fold symmetric diffraction spot (Figure 4d) with the diffraction intensity ratio of the outer (2-1-10) peak over the inner (1-110) peak is approximately 1 (Figure 4h), indicating the single-crystalline nature of the monolayer graphene seeding layer. Meanwhile, SAED patterns (Figure 4e-g) from the marked red regions in Figure 4a show the similar diffraction patterns as the seeding monolayer graphene, with the diffraction intensity ratio of the outer peak over the inner peak is approximately 2 (Figure 4i-k), which evidenced the AB stacking order of the as-grown bilayer film. The same crystalline orientation of the bilayer graphene confirms that the second layer graphene were epitaxial grown on the transferred single crystal monolayer graphene as expected[47, 48]. As schematically illustrated in Figure 4c, multiple BLG nuclei form with perfect rotational registry on the single-crystal seed and subsequently coalesce into a continuous single-crystal layer without grain-boundary defects, even at high nucleation density.

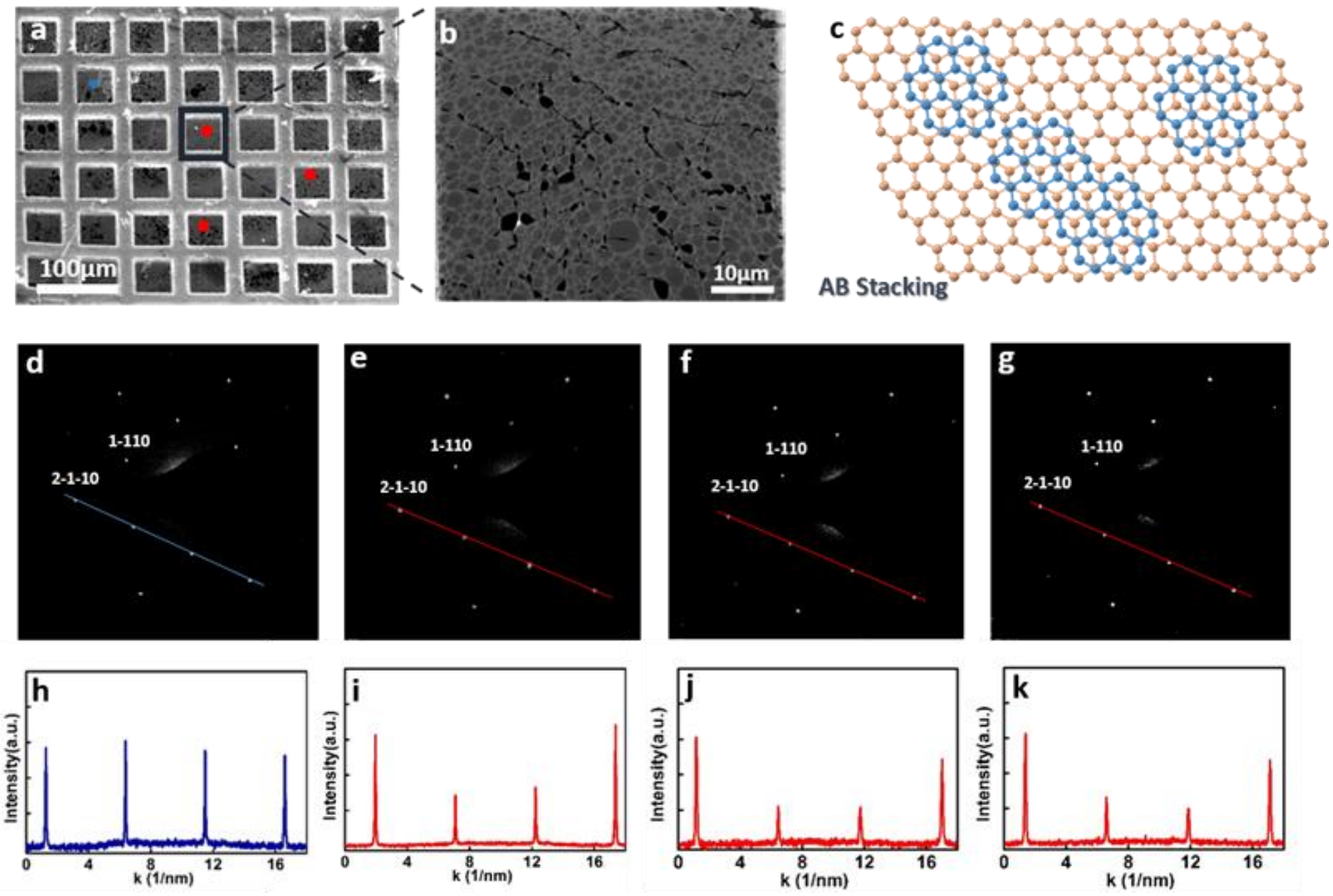

**Figure 4.** (a) SEM image of the as-grown BLG transferred onto the TEM grid. The blue and red circles denote the single crystal graphene substrate and BLG area, respectively. (b) Higher magnification SEM images of the corresponding black square marked in (a). (c) A schematic drawing of the synthesis of AB stacked bilayer graphene on the single crystal graphene seeding layer. (d) SAED pattern of the monolayer graphene in the blue circle region marked in (a). (e-f) SAED patterns of the as-grown BLG in the red circle regions marked in (a), all the three regions have the same orientation as the seeding layer. (h-k) Intensity profiles along the lines in (d-g), respectively.

To further evaluate the electronic quality of the BLG films, we fabricated the dual-gate graphene field-effect transistors (FETs). The schematic structure of the device is shown in Figure 5a (See Experimental Section for more details). Because Hafnium oxide ($HfO_2$) film couldn't be directly grown on graphene layers by atomic layer deposition (ALD)[49], an ultrathin Yttrium oxide ($Y_2O_3$) film (5 nm) was first deposited, followed by 20 nm $HfO_2$ to provide ideal gate dielectric for graphene-based FETs[50, 51]. Figure 5b shows the representative optical image of a dual-gate BLG device. The resistance $R$ versus the back-gated voltage ($V_{BG}$) ($R$-$V_{BG}$) plot in Figure 5c shows typical ambipolar characteristics expected for the graphene devices and the back-gate mobility is extracted to be 2297 $cm^2/Vs$ based on the fitting method reported previously[49]. Furthermore, we swept the voltage of top gate ($V_{TG}$) from -4 V to 4 V at different back gate voltages ($V_{BG}$) from -80 V to 80 V applied on the silicon substrate. The two-dimensional plot of the device resistance $R$ versus $V_{TG}$ and $V_{BG}$ show that the highest resistance is achieved in the highest displacement field region (top middle left and bottom middle right) (Figure 5d). This can be more evidently illustrated in a series of plots of $R$-$V_{TG}$ at different $V_{BG}$ in Figure 5e, confirming the AB stacking nature of the bilayer graphene[52]. For each trace, a resistance maximum, corresponding to the charge neutrality point (CNP), was seen. The CNP resistance is adjusted by tuning the vertical electric field which proves the formation of a tunable band structure[53, 54]. This was confirmed by the observation of an enhancement in the current on/off ratio. Figure 5e also shows that the $I_{on}/I_{off}$ values obtained in all measurements were about 1.66 and 6.46 for $V_{BG}$ = 80 V and − 80 V, respectively. Figure 5f shows that $V_{Dirac}$ and $V_{BG}$ are linearly related with a slope of about -0.017, agrees well with the expected value of -$\varepsilon_{BG}\, d_{TGall} / \varepsilon_{TGall}\, d_{BG}$ = -0.027, where $\varepsilon$ and $d$ correspond to the dielectric constant and thickness of the top gate ($Y_2O_3$: $d_1$ = 5 nm, $\varepsilon_1$ = 12; $HfO_2$: $d_2$ = 20 nm, $\varepsilon_2$ = 12, $d_{TGall}$ = 25 nm, $\varepsilon_{TGall}$ = 12) and the bottom gate ($SiO_2$: $d_{BG}$ = 300 nm, $\varepsilon_{BG}$ = 3.9). The carrier mobility of the top-gate BLG on $SiO_2$/Si substrates is 2297 $cm^2$/Vs, which is comparable to those of CVD grown bilayer graphene on metal substrates[11]. These results clearly demonstrate that the bilayer graphene obtained by the direct epitaxial growth method on Si wafer is of high-quality, and the proposed strategy provides a rational route to synthesis high-quality AB-stacked bilayer graphene directly on the dielectric substrates.

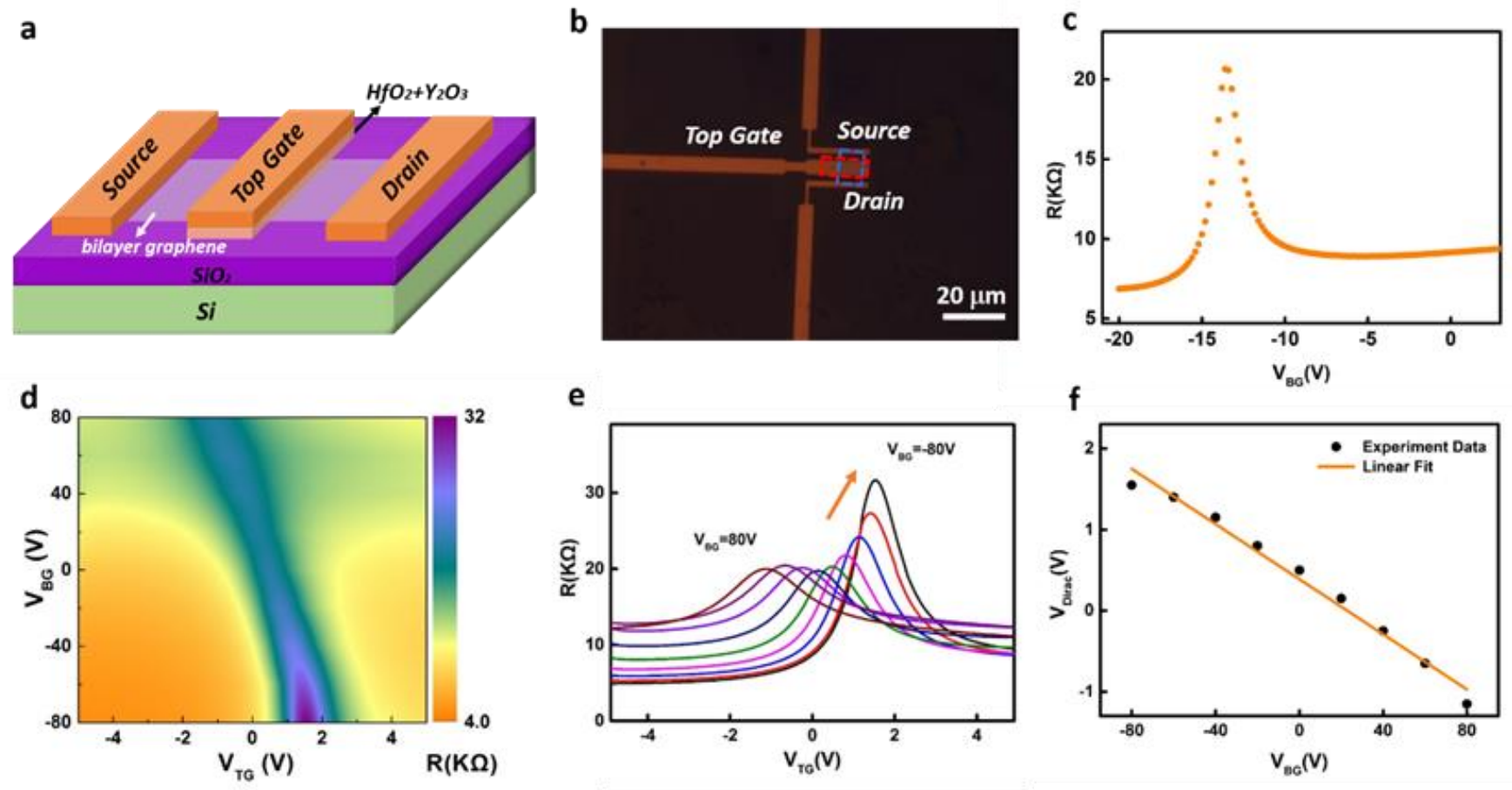

**Figure 5.** (a-b) Schematic structure diagram (a) and optical microscope image (b) of the BLG device with dual-gate applied for the transfer characteristics measurements. The BLG sheet and dielectric layer are marked by the blue and red dashed frame, respectively. (c) Plots of device resistance ($R$) versus back-gate voltage ($V_{BG}$). (d) Two-dimensional plot of resistance as functions of both top-gate voltage ($V_{TG}$) and $V_{BG}$ of the dual-gate BLG device. (e) Resistance measured as a function of the top-gate voltage ($V_{TG}$) at a range of fixed back-gate voltages ($V_{BG}$) from − 80 V to + 80 V. The traces were taken at 20 V steps in the back-gate voltage. (f) Linear relation between the top-

gate neutral points and the back-gate voltages.

## 4. Conclusions

In conclusion, we have proposed a new method to grow single-crystalline bilayer graphene directly on various dielectric substrates. The quality and stacking order of bilayer graphene were confirmed by Raman spectroscopy, TEM diffraction patterns and dual gated FET measurements. Transport measurements show the bilayer AB-stacked graphene has a typical tunable transport band gap. The measured mobility of the bilayer graphene was 2297 $cm^2$/ Vs. This Cu NP assisted graphene seeding growth technique can be beneficial for producing AB stacked bilayer graphene with high quality directly on insulting substrate, which is significant for the further development of functional graphene devices.

**Acknowledgements:**

We appreciate the support from the National Key Research and Development Program of China (Grant No. 2024YFA1409700), and the Fundamental Research Funds for the Central Universities and the Research Funds of Renmin University of China (Grant No. 21XNLG26). Zuoquan Tan was supported by the Outstanding Innovative Talents Cultivation Funded Programs 2024 of Renmin University of China.

**Supplementary Material:** Supplementary material is available in supporting information.